\documentstyle [twoside,12pt]{article}
\newcommand{\nc}{\newcommand}
\nc{\ra}{\rightarrow}
\nc{\la}{\lambda} \nc{\al}{\alpha}
\nc{\th}{\theta}  \nc{\be}{\beta}
\nc{\ga}{\gamma}  \nc{\Ga}{\Gamma}
\nc{\de}{\delta}  \nc{\De}{\Delta}
\nc{\si}{\sigma}
\nc{\om}{\omega}  \nc{\Om}{\Omega}
\nc{\beq}{\begin{equation}}
\nc{\eeq}{\end{equation}}
\nc{\beqa}{\begin{eqnarray}}
\nc{\eeqa}{\end{eqnarray}} \nc{\nnb}{\nonumber}

\title{ \large{\bf On a possible breaking of global N=2 supersymmetry in
non-linear $\si$ models on compact K\"ahler target spaces}}
\author{Guy Bonneau\thanks
{\noindent Laboratoire de Physique Th\'eorique et des Hautes Energies,
 Unit\'e associ\'ee au CNRS URA 280,~Universit\'e Paris 7,
 2 Place Jussieu, 75251 Paris Cedex 05; e-mail adress :
bonneau@lpthe.jussieu.fr.}}
\begin{document}
\maketitle
\begin{abstract}
\noindent We analyse with the algebraic, regularisation independent,
cohomological B.R.S. methods, the renormalisability of torsionless N=2
supersymmetric non-linear $\si$ models built on compact K\"ahler spaces.
Surprisingly enough with respect to the common wisdom, we obtain an anomaly
candidate, when the Hodge number $h^{3,0}$ of the target space manifold is
different from zero : this occurs in particular in the Calabi-Yau case. On the
contrary, in the compact homogeneous K\"ahler case, the anomaly candidate
disappears.
\end{abstract}
\vspace{2cm}
{\bf Contribution to the $XI^{th}$ International Congress of Mathematical
Physics, Paris} \newline and talk given to the Satellite Conference :{\bf``New
problems in the General Theory of Fields and Particles"}

\vfill
\hfill  May 1994

\section{Introduction}
Supersymmetric non-linear $\si$  models in two space time dimensions have been
considered for many years to describe the vacuum state of superstrings
\cite{1}\cite{1a}. In particular Calabi-Yau spaces, {\it i.e.} 6 dimensional
compact K\"ahler Ricci-flat manifolds \cite {2}, appear as good candidates in
the compactification of the 10 dimensional superstring to 4 dimensional flat
Minkowski space ; the conformal invariance of the 2.d, N = 2 supersymmetric
non-linear $\si$  model (the fields of which are coordinates on this compact
manifold) is expected to hold to all orders of perturbation theory \cite {3}.

However explicit calculations to 4 or 5 loops \cite {4} and, afterwards,
general arguments \cite{5} show that the $\be$ functions may not vanish.
But, as argued in a recent review \cite{6}, at least two problems obscure these
analyses : first, the fact that the quantum theory is not sufficiently
constrained by the K\"ahler Ricci-flatness requirement ; second, the use of
``dimensional reduction" \cite{62} or of harmonic superspace formalism
\cite{61} \footnote{\ The regularisation through dimensional reduction suffers
from algebric inconsistencies and the quantization in harmonic superspace does
not rely on firm basis, due to the presence of non-local singularities ( in the
harmonic superspace) \cite{63}.} in actual explicit calculations and general
arguments. Then, we prefer to analyse these models using the B.R.S., algebraic,
regularisation free cohomological methods.

So we adress ourselves to the question of the all-order renormalisability of N
= 2 supersymmetric non linear $\si$   models in two space time dimensions.

Our main result is that, surprisingly enough with respect to the common wisdom
\footnote{\ Notice also that recent works of Brandt \cite{8} and Dixon \cite{9}
show the existence of new non-trivial cohomologies in supersymmetric
theories.}, there exists a possible anomaly for \underline{global}
supersymmetry in 2 space-time dimensions, at least for \underline{torsionless}
compact K\"ahler Ricci-flat manifolds ({\it i.e.} special N=2 supersymmetric
models)\cite{bo}.

\section {The classical theory and the Slavnov operator}

We use N = 1 superfields \footnote{\ The quantization with N = 1 superfields
was put on firm basis by Piguet and Rouet \cite{10} who proved in particular
the Quantum Action Principle in that context. Moreover, in \cite{13}a) we show
the renormalisability of N=1 supersymmetric non-linear $\si$ models using
component fields : this justifies the use of N=1 superfields for the present
analysis of extended supersymmetry. \newline Notice also that we use light-cone
coordinates.  } $\phi^i (x^+,x^-,\th^+,\th^-)$  and, in the  \underline{absence
of torsion}, the most general N = 1 invariant action is :
\beq\label{a1}
S^{inv.} = \int d^2x d^2\th g_{ij}[\phi]D_+\phi^iD_-\phi^j
\eeq
where the supersymmetric covariant derivatives $$ D_{\pm} =
\frac{\partial}{\partial \th^{\pm}} + i\th^{\pm} \frac{\partial}{\partial
x^{\pm}}  $$ satisfy $$ \{ D_+ , D_- \} = 0 \ \ \ ,\ \ D^2_{\pm} =
i\frac{\partial}{\partial x^{\pm}} \equiv i\partial _{\pm} \ \ .$$

\noindent The tensor $ g_{ij}[\phi]$ is interpreted as a metric tensor on a
Riemannian manifold $\cal M$ \footnote {\ Here, contrarily to our previous work
where the manifold $\cal M$ was supposed to be an homogeneous space \cite {7},
we consider renormalisability ``{\it \`a la Friedan} " \cite {100}, {\it i.e.}
in the space of metrics, and analyse only the possibility of maintaining to all
orders the N=2 supersymmetry. As explained in \cite {6}, in order to define
unambiguously the classical action, one should add extra properties. }. As is
by now well known \cite {11}, N = 2 supersymmetry needs $\cal M$ to be a 2n
dimensional K\"ahler manifold.
The second supersymmetry transformation writes :
\beq\label{a2}
\de\phi^i \ =\ J^i_j[\phi](\epsilon^+ D_+\phi^j + \epsilon^- D_-\phi^j)\ .
\eeq
where $J^i_j$ is a covariantly constant integrable complex structure.
In the B.R.S. approach, the supersymmetry parameters $\epsilon^{\pm}$ are
promoted to constant, commuting Faddeev-Popov parameters $d^{\pm}$   and an
anticommuting classical source $\eta_i$  for the non-linear field
transformation (\ref{a2}) is introduced in the classical action \footnote {\ In
the absence of torsion, there is a parity invariance $$ + \ra -, d^2x \ra d^2x,
d^2\th \ra -d^2\th, \phi^i \ra \phi^i, \eta_i \ra -\eta_i \ .$$ Moreover, the
canonical dimensions of $[ d^2x d^2\th],\ [\phi^i],\ [d^{\pm}],\ [D_{\pm}],\
[\eta_i]$ are -1, 0, -1/2, +1/2, +1 respectively and the Faddeev-Popov
assignments + 1 for $d^{\pm}$, -1 for $\eta_i$, 0 for the other quantities.}:
\beq\label{a3}
\Ga^{class.} = \int d^2x d^2\th \left\{ g_{ij}[\phi]D_+\phi^iD_-\phi^j + \eta_i
J^i_j[\phi](d^+ D_+\phi^j + d^- D_-\phi^j) \right\}\ .
\eeq
For simplicity, no mass term is added here as we are only interested in U.V.
properties.
The non linear Slavnov operator is defined by
$$S\Ga \equiv \int d^2x d^2\th \frac {\de\Ga}{\de\eta_i(x,\th)}\frac
{\de\Ga}{\de\phi^i(x,\th)} $$
and we find $$S\Ga^{class.} = (d^+)^2\int d^2x d^2\th \eta_k
i\partial_{+}\phi^k + (d^-)^2\int d^2x d^2\th \eta_k i\partial_{-}\phi^k $$
in accordance with the supersymmetry algebra.

As is by now well known (for example see \cite{6} or \cite{7}), in the absence
of a consistent regularisation that respects all the symmetries of the theory,
the quantum analysis directly depends on the cohomology of the nihilpotent
linearized Slavnov operator :
\beqa\label{a4}
S_L & = & \int d^2x d^2\th \left[\frac {\de\Ga^{class.}}{\de\eta_i(x,\th)}\frac
{\de}{\de\phi^i(x,\th)} + \frac {\de\Ga^{class.}}{\de\phi^i(x,\th)}\frac
{\de}{\de\eta_i(x,\th)} \right] \nnb\\
S_L^2 &=& 0
\eeqa
in the Faddeev-Popov charge +1 sector [absence of anomalies for the N = 2
supersymmetry] and charge 0 sector [number of physical parameters and stability
of the classical action through renormalization]. Notice that the Slavnov
coboundaries $S_L \int d^2x d^2\th \eta_i W^i[\phi]$ correspond to field and
source reparametrisations :
\beq\label{a40}
\phi^i \ra \phi^i + \la W^i[\phi] \ \ \ ,\ \ \eta_i \ra \eta_i - \la \eta_k
W^k_{,i}[\phi] \ ,
\eeq
where $W^i[\phi]$ is an arbitrary function of the fields $\phi(x,\th)$ and a
comma indicates a derivative with respect to the field $\phi^i$.

Due to the highly non-linear character of $S_L$ (equ.(\ref{a4})), it is
convenient to use a ``filtration" (\cite{14},\cite{7}) with respect to the
number of fields $\phi^i(x,\th)$ and their derivatives. As it does not change
this number, the nihilpotent lowest order part of $S_L$, $S_L^0$, will play a
special role :
\beqa\label{a5}
S_L &=& S_L^0 + S_L^1 + S_L^2 +... \equiv S_L^0 + S_L^r \ ,\ \ (S_L^0)^2 =
(S_L^r)^2 + S_L^0 S_L^r + S_L^r S_L^0 = 0 \nnb\\
S_L^0 &=&  \int d^2x d^2\th J^i_j(0) \left\{ (d^+ D_+\phi^j + d^-
D_-\phi^j)\frac{\de}{\de\phi^i} + (d^+ D_+\eta_i + d^-
D_-\eta_i)\frac{\de}{\de\eta_j}\right\}\ .
\eeqa

\noindent As explained in refs.\cite{7} and \cite{13}a), when $S_L^0$ has no
cohomology in the Faddeev-Popov positively charged sectors, the cohomology of
the complete $S_L$ operator in the Faddeev-Popov sectors of charge 0 and +1 is
isomorphic to a subspace of the one of  $S_L^0$ in the same sectors.

Then, we first analyse the possible cohomology of the Slavnov operator in the
anomaly sector : indeed, if there exists a non-trivial cohomology, this is the
death of the theory (for a complete analysis we refer to \cite{bo} and
\cite{13}b)).

\section{A candidate for an anomaly}

According to the spectral sequence method, we first analyse the possible
cohomology of $S_L^0$ in the anomaly sector.

\subsection{$S_L^0$ cohomology}
The most general dimension zero integrated local polynomial in the
Faddeev-Popov parameters, fields and their derivatives, of Faddeev-Popov +1
charge writes :
\beqa\label{a10}
\De_{[+1]} &=& \int d^2x d^2\th \{ t^{[ijk]}(d^+)^2(d^-)^2\eta_i\eta_j\eta_k
\nnb\\
&+& d^+d^-[\eta_i\eta_j t^{[ij]}_{1\; n}(d^+D_+\phi^n - d^-D_-\phi^n) + \eta_n
t^n_{2\; [ij]}D_+\phi^iD_-\phi^j ]\nnb\\
&+& d^+d^- \eta_i s^{(ij)}_1 (d^+D_+\eta_j - d^-D_-\eta_j) \nnb\\
&+& (d^+)^2(\eta_n t^n_{3\; [ij]}D_+\phi^iD_+\phi^j +
D_+\eta_iD_+\phi^jt^i_{4\;j})\nnb\\
&+& (d^-)^2(\eta_n t^n_{3\; [ij]}D_-\phi^iD_-\phi^j +
D_-\eta_iD_-\phi^jt^i_{4\;j})\nnb\\
&+& d^+(\tilde {t}_{[ij]n}D_+\phi^iD_+\phi^jD_-\phi^n + s_{2\;
(ij)}D_-D_+\phi^iD_+\phi^j)\nnb\\
&-& d^-(\tilde {t}_{[ij]n}D_-\phi^iD_-\phi^jD_+\phi^n + s_{2\;
(ij)}D_+D_-\phi^iD_-\phi^j)\}
\eeqa
where, due to the anticommuting properties of $\eta_i$ and $D_{\pm}\phi^i$ and
to the integration by parts freedom, the tensors $t^{[ijk]}$, $t^{[ij]}_{1\;
n}$, $t^n_{2\; [ij]}$, $t^n_{3\; [ij]}$, $\tilde {t}_{[ij]n}$ are antisymmetric
in i, j, k, and $s^{(ij)}_1$, $s_{2\; (ij)}$ symmetric in i, j.

The analysis of the cocycle condition $S_L^0 \De_{[+1]} = 0$ leads to :
\beqa\label{a91}
\De_{[+1]} & = &   \De_{[+1]}^{an.}[t^{[ijk]}(\phi)] + S_L^0 \De_{[0]}[{\rm
arbitrary\ }t_{ij}(\phi)\ {\rm and\ } U^i_j(\phi)] \\
{\rm where}\ \ \De_{[0]} & = & \int d^2x d^2\th \left\{
t_{ij}[\phi]D_+\phi^iD_-\phi^j + \eta_i U^i_j[\phi](d^+ D_+\phi^j + d^-
D_-\phi^j) \right\} \nnb
\eeqa
and where the antisymmetric tensor $t^{[ijk]}(\phi)$ which occurs in the
anomalous part
\beq\label{a92}
\De_{[+1]}^{an.} = \int d^2x d^2\th
t^{[ijk]}(\phi)(d^+)^2(d^-)^2\eta_i\eta_j\eta_k
\eeq
is constrained by :
\beqa\label{a11}
a)& \ \ J^i_n(0)t^{[njk]} \ \ \ \ {\rm is\ \ i,\ j,\ k\ \ antisymmetric},\nnb\\
b)&  \ \ J^i_n(0)t^{[njk]}_{,m} = J^n_m(0)t^{[ijk]}_{,n}
\eeqa

\noindent These conditions, when expressed in a coordinate system adapted to
the complex structure  $J^i_j[\phi]$ (i $\equiv (\al,\bar{\al})$ :
$J^{\al}_{\be} = i \de^{\al}_{\be} , J^{\bar{\al}}_{\bar{\be}} = -i
\de^{\al}_{\be}, J^{\bar{\al}}_{\be} = J^{\al}_{\bar{\be}} = 0 $) mean that the
tensor  $t^{[ijk]}$ is a pure contravariant skewsymmetric analytic tensor ({\it
i.e.} $t^{[\al\be\ga]} = t^{[\al\be\ga]}(\phi^{\de})$ ,
$t^{[\bar{\al}\bar{\be}\bar{\ga}]}  =
t^{[\bar{\al}\bar{\be}\bar{\ga}]}(\bar{\phi}^{\bar{\de}}))$ , the other
components vanish). In particular, due to the vanishing of
$t^{[\al\be\bar{\ga}]}$,  such tensor cannot be a candidate for a torsion
tensor on a K\"ahler manifold \cite{151}.

Consider the covariant tensor
$$t_{[\al\be\ga]} =
g_{\al\bar{\al}}g_{\be\bar{\be}}g_{\ga\bar{\ga}}
t^{[\bar{\al}\bar{\be}\bar{\ga}]}\ .$$
It satisfies $\nabla _{\de}t_{[\al\be\ga]} = 0\ .$ Then the (3-0) form
$$\om ' = \frac{1}{3!}t_{[\al\be\ga]}d\phi^{\al}\wedge d\phi^{\be} \wedge
d\phi^{\ga}$$ which satisfies $d'\om' = 0$, may be shown to be harmonic as
$\cal{M}$ is a \underline{compact} manifold
\footnote{\ In this K\"ahlerian case, one firstly obtains from $\nabla
_{\de}t_{[\al\be\ga]} = 0\ ,\ \ \ \triangle t_{[\al\be\ga]} = g^{\de\bar{\de}}
\nabla _{\de}\nabla _{\bar{\de}}t_{[\al\be\ga]} -[R^{\de}_{\al}t_{[\de\be\ga]}
+ \ {\rm perms.}\ ]\ $ ; on another hand, the Ricci identity gives
$g^{\de\bar{\de}} \nabla _{\de}\nabla _{\bar{\de}}t_{[\al\be\ga]} =
-[R^{\de}_{\al}t_{[\de\be\ga]} + \ {\rm perms.}\ ].$
So $ \triangle t_{[\al\be\ga]} = 2g^{\de\bar{\de}} \nabla _{\de}\nabla
_{\bar{\de}}t_{[\al\be\ga]}$. Now, the manifold being compact, one may compute
:
$$ (d\om ',d\om ') + (\de\om ',\de\om ') = (\om ',(d\de+\de d)\om ') = (\om
',\triangle\om ') = $$
$$= \int_{\cal{M}} d\si 2t^{[\al\be\ga]}g^{\de\bar{\de}} \nabla _{\de}\nabla
_{\bar{\de}}t_{[\al\be\ga]} = \int_{\cal{M}} d\si 2g^{\de\bar{\de}} \{\nabla
_{\de}\nabla _{\bar{\de}}(t^{[\al\be\ga]}t_{[\al\be\ga]}) - \nabla
_{\de}t^{[\al\be\ga]}\nabla _{\bar{\de}}t_{[\al\be\ga]}\} = 0 - 2(d\om ',d\om
')$$
$$\Rightarrow \ (\de\om ',\de\om ') + 3(d\om ',d\om ') = 0
\ \Rightarrow \ \de\om ' = d\om '= \triangle \om' = 0\ . \ \ \ Q.E.D.$$ }
(\cite{16},\cite{13}b)). It is known that the number of such forms is given by
the Hodge number $h^{3,0}$ : then this number determines an upper bound for the
dimension of the cohomology space of $S_L$ in the anomaly sector.

As a first result, this proves that if the manifold $\cal M$ has a complex
dimension smaller than 3, there is no anomaly candidate. Another special case
is the compact K\"ahler homogeneous one ( N=2 supersymmetric extension of our
previous work on the bosonic case \cite{7}) : in such a case the Ricci tensor
is positive definite \cite{77} which forbids (\cite{16},\cite{13}b)) the
existence of such analytic tensor $t^{[\al\be\ga]}(\phi^{\de})$. As a
consequence, the cohomology of $S_L^0$ - and then of $S_L$ - vanishes in the
anomaly sector (for details, see ref.\cite{13}b)).

On the contrary, when $h^{3,0}\neq 0$, we have a candidate for an anomaly. We
shall now discuss the possible non-trivial cohomology of the complete $S_L
\equiv S_L^0 + S_L^r$ operator, still in the Faddeev-Popov charge +1 sector.

\subsection{$S_L$ cohomology}

Starting from the $S_L^0$ cohomology (\ref{a92}), we were able to construct the
$S_L$ cohomology in the same Faddeev-Popov sector \footnote{\ In this compact
case, due to  the harmonicity of the (3,0) form $\om'$, one gets from $d\om' =
d"\om' =0$ the vanishing of $\nabla _{\bar{\de}}t_{[\al\be\ga]}$, which means
that $t_{ijk}$ is a pure covariant skewsymmetric analytic tensor.}
(\cite{13}b)) :
\beqa\label{a12}
\De_{[+1]}^{an.} & = & \int d^2x d^2\th t^{[ijk]}[\phi]
\{(d^+)^2(d^-)^2\eta_i\eta_j\eta_k \nnb\\
& - & {3\over 2}d^+d^-(\eta_i\eta_j J_{kn}(d^+D_+\phi^n - d^-D_-\phi^n) +
2\eta_i J_{jn} J_{km}D_+\phi^nD_-\phi^m )\nnb\\
& + & {3\over 4}  J_{in} J_{jm} J_{kl}(d^+D_+\phi^n D_+\phi^m D_-\phi^l -
d^-D_-\phi^n D_-\phi^m D_+\phi^l) \}
\eeqa

\noindent As a consequence, if at a given pertubative order this anomaly
appears with a non zero coefficient
$$S_L\Ga|_{p^{th} order} = a (\hbar)^p \De_{[+1]}^{an.}, \ \ a \neq 0 $$
the N = 2 supersymmetry is broken as $\De_{[+1]}^{an.}$ cannot be reabsorbed
(being a  cohomology element, it is not a $S_L \tilde{\De}_{[0]}$ ) and, {\it a
priori}, we are no longer able to analyse the structure of the U.V. divergences
at the next perturbative order, which is the death of the theory.
When the target space is a Ricci-flat one, such tensor may exist : Calabi-Yau
manifolds (3 complex dimensional case) where $h^{3,0} = 1 $ \cite{2} are
interesting examples \footnote{\ As $\det\|g\| = 1$, a representative of
$t^{[\al\be\ga]}$ is the constant skewsymmetric tensor
$\epsilon^{[\al\be\ga]}$( with $\epsilon^{123} = +1$).}
due to their possible relevance for superstring theories. Of course, as no
explicit metric is at hand, one can hardly compute the anomaly coefficient.

\section {Concluding remarks}

We have analysed the cohomology of the B.R.S. operator associated to N = 2
supersymmetry in a N = 1 superfield formalism. We have found an anomaly
candidate for torsionless models built on compact K\"ahler target spaces with a
non vanishing Hodge number  $h^{3,0}$. This anomaly in \underline{global}
extended supersymmetry is a surprise with respect to common wisdom \cite{18} (
but see other unexpected cohomologies  in supersymmetric theories, in the
recent works of Brandt \cite{8} and Dixon \cite{9}).

Of course, our analysis casts some doubts on the validity of the previous
claims on U.V. properties of N=2 supersymmetric non linear $\si$ models :
there, the possible occurence at 4-loops order of (infinite) counterterms
non-vanishing on-shell, even for K\"ahler Ricci-flat manifolds, did not ``
disturb" the complex structure ; on the other hand, we have found a possible ``
instability" of the second supersymmetry, which confirms that there are some
difficulties in the regularisation of supersymmetry by dimensional reduction
assumed as well in explicit perturbative calculations \cite{4} than in
finiteness ``proofs" \cite{3} or higher order counterterms analysis \cite{5}.
We would like to emphasize the difference between Faddeev-Popov 0 charge
cohomology which describes the stability of the classical action against
radiative corrections ( the usual ``infinite" counterterms) and which offers no
surprise(see \cite{13}b) for a detailed analysis), and the anomaly sector which
describes the ``stability" of the symmetry ( the finite renormalisations which
are needed, in presence of a regularisation that does not respect the
symmetries of the theory, to restore the Ward identities) : of course, when at
a given perturbative order the Slavnov (or Ward) identities are spoiled, at the
next order, the analysis of the structure of the divergences is no longer under
control.
In particular, the Calabi-Yau uniqueness theorem for the metric \cite{19}
invoked in some finiteness ``proofs" \cite{3} supposes that one stays in the
same cohomology class for the K\"ahler form, a fact which is not certain in the
absence of a regularisation that respects the N=2 supersymmetry (the possible
anomaly we found expresses the impossibility to find a regularisation that
respects all the symmetries of these theories).

We emphasize the fact that the present work relies heavily on perturbative
analysis, especially through the use of the Quantum Action Principle in order
to analyse the possible breakings of Slavnov identities. It may well happen
that the coefficient of the anomaly candidate vanishes at any finite order of
perturbation theory \footnote{\ An argument based on the universality of the
coefficient at any finite order of perturbation theory and its vanishing for a
special class of Calabi-Yau manifolds corresponding to orbifolds of tori has
been given to the author by the referee of \cite{bo}.}, the existence of a
non-trivial supersymmetry cohomology leaving open the interesting possibility
of a non-perturbative breaking of N=2 supersymmetry.

Finally, when the manifold is a compact \underline{homogeneous} K\"ahler space,
the anomaly candidate disappears as expected. Moreover, we have also been able
to prove that, if one enforces N=4 supersymmetry (\underline{HyperK\"ahler
manifolds}), the anomaly vanishes (\cite{13}b)).

\bibliographystyle{plain}

\end{document}